\documentclass[%
superscriptaddress,
twocolumn,
 amsmath,amssymb,
 aps,
 prx,
]{revtex4-2}
\usepackage{graphicx}
\usepackage{dcolumn}
\usepackage{bm}
\usepackage{comment}
\usepackage{hyperref}
\usepackage{amsmath}
\usepackage{color}

\newcommand\Peclet{\mbox{\textit{Pe}}}
\newcommand{\red}[1]{{\leavevmode\color{black}#1}}

\begin{document}

\title{Active Gel Theory for Cell Migration With Two Myosin Isoforms}

\author{Nils O. Winkler}
\affiliation{Institute for Theoretical Physics and Bioquant, Heidelberg University, 69120 Heidelberg, Germany}
\author{Oliver~M.~Drozdowski}
\affiliation{Institute for Theoretical Physics and Bioquant, Heidelberg University, 69120 Heidelberg, Germany}
\affiliation{Max Planck School Matter to Life, Heidelberg University, 69120 Heidelberg, Germany}
\author{Falko~Ziebert}%
\affiliation{Institute for Theoretical Physics and Bioquant, Heidelberg University, 69120 Heidelberg, Germany}
\author{Ulrich~S.~Schwarz}%
 \email{Corresponding author: schwarz@thphys.uni-heidelberg.de}
\affiliation{Institute for Theoretical Physics and Bioquant, Heidelberg University, 69120 Heidelberg, Germany}
\affiliation{Max Planck School Matter to Life, Heidelberg University, 69120 Heidelberg, Germany}
\date{\today}

\begin{abstract}
Myosin II molecular motors slide actin filaments relatively to each other 
and are essential for force generation, motility and mechanosensing in animal cells.
For non-muscle cells, evolution has resulted in three different isoforms, which have
different properties concerning the motor cycle 
and also occur in different abundances in the cells, but their respective
biological and physical roles are not fully understood. Here we use
active gel theory to demonstrate the complementary roles of isoforms A and B
for cell migration. We first show that our model can be derived
both from coarse-graining kinetic equations and from nonequilibrium
thermodynamics as the macroscopic limit of a two-component Tonks gas.
We then parametrize the model and show that 
motile solutions exist, in which the more abundant and more dynamic isoform A is
localized to the front and the stronger isoform B to the rear, 
in agreement with experiments. 
\red{Exploring parameter space beyond the isoform parameters
typical for animal cells, we also find cell oscillations in length and velocity, which might
be realized for genetically engineered systems.
We also describe an analytical solution for the stiff limit,
which then is used to calculate a state diagram,
and the effect of actin polymerization at the boundaries, that leads to an imperfect pitchfork bifurcation.}
Our findings highlight the importance of including isoform-specific molecular 
details to describe whole cell behavior.
\end{abstract}


\maketitle

\section{Introduction}

Cell migration is an essential process for any living organism. In multicellular animals like humans,
it is important mainly for embryonic development \cite{embryo}, wound healing \cite{trepat2009physical} and the immune response \cite{nourshargh2010breaching}, but also for the spread of cancer \cite{cancer}. 
It is strongly linked to the underlying molecular processes, because cells have to
generate force and movement that is converted into consistent migration on the cell scale.
This function is provided by the cytoskeleton, a collection of polymer networks.
Cells can push their envelope forward by polymerization of these filaments.
They also can generate pulling forces by sliding them relative to each other
with molecular motors. For animal cells, pushing and pulling forces
are generated mainly by actin filaments and myosin II 
molecular motors. In cell migration, they go hand-in-hand, 
because the actin networks generated by polymerization 
are pulled backward by myosin II motors.
The resulting retrograde flow is
then transmitted to the substrate by adhesions
and thus generates forward motion of the cell \cite{blanchoin2014actin}.
Migration requires a cyclic process that
converts energy provided in the form of ATP into force and movement.
Moreover, initiation of motility requires a symmetry
break between front and back, which can be either
spontaneous or driven by some external cue \cite{VERKHOVSKY199911,Ridley_2003,Ziebert2012}. 
Usually this break of symmetry is related
to activation of the Rac/Cdc42 and RhoA pathways, which for quantitative experiments can be controlled by optogenetics \cite{Hadjitheodorou_2021_NatComm_Reorientation_migrating_neutrophils,Valon_2017_Nature_Optogenetic_control_cellular_forces,hennig_sciadv_stick_slip_optogenetic_1d}. 

\begin{figure}[b!]
    \centering
    \includegraphics[width=.9\linewidth]{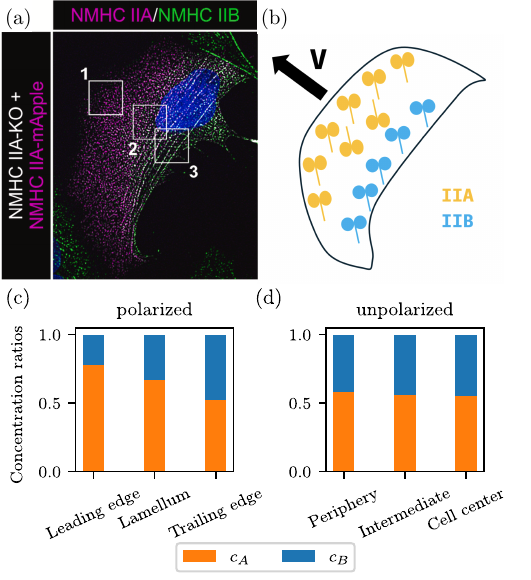}
    \caption{NMII isoform distributions in cells. 
    (a) Microscopy image of a U2OS cell
    with exogenous NMIIA (magenta) and endogenous NMIIB (green). Modified 
    from \cite{bastmeyer_graphic}.
    (b) Schematic sketch of a cell indicating NMII isoform distribution and net cell movement.
    (c),(d) Experimental concentration ratios of NMIIA and NMIIB within different cell compartments (c: along the axis of polarization, d: unpolarized cell). Adapted from \cite{bastmeyer_graphic}. 
    While both isoforms are equally distributed in an unpolarized cell, in a moving cell 
    NMIIA and NMIIB enrich at the leading and trailing edges, respectively.}
    \label{fig:fig1}
\end{figure}

In non-muscle animal cells contractile stress for
cell motility is produced by motor proteins of the non-muscle myosin II (NMII) family \cite{manzanares2009}. Evolution
has resulted in three isoforms, namely NMIIA, NMIIB and NMIIC \cite{weissenbruch2024actomyosin}. 
After activation of the Rho-pathway, the myosin motors are assembled into minifilaments that typically mix different isoforms \cite{beach2014nonmuscle,weissenbruch2024actomyosin}. These minifilaments
then bind to the actin cytoskeleton, generate stress and at the same time are advected with the resulting cytoskeletal flow. 
In the case of asymmetric contractile stress distributions, the cell is polarized and starts migrating. 
While NMIIA is most abundant and propels loads the quickest along actin filaments, NMIIB is the strongest due to its higher duty ratio; NMIIC seems to have little relevance for cell migration, also due to its low abundance \cite{eLife,bastmeyer_graphic}. 
Recent experiments have shown that in a migrating cell the isoforms A and B are not distributed equally \cite{beach2014nonmuscle,bastmeyer_graphic}.
While the more abundant and more dynamic isoform A localizes more at the leading edge, the stronger isoform B accumulates
towards the trailing edge (cf.~Fig.~\ref{fig:fig1}). 

In the following, we will define and analyze a model that considers 
the complementary roles played by the two dominant isoforms A and B.
We use active gel theory, which models the contractile actomyosin network as an active viscous material
and thus is a natural
approach to describe flowing actin networks as the physical basis of cell crawling 
\cite{Kruse_2004_PRL,Kruse2004_paradigm}.
In principle, active gel theory can describe 
the effects of both actin polymerization and myosin-based
contractility, but seminal work has shown that
contractility alone is sufficient to explain cell migration
\cite{RechoPRL,Recho_motility_initiation_and_arrest,DrozdowskiPRE2021}. 
While this minimal model leads to a supercritical pitchfork bifurcation
towards a motile state, more recently
it was shown that a subcritical 
pitchfork bifurcation and bistability can be obtained
when including a non-linearity in the 
diffusion behavior, which can be derived
from the physics of a van der Waals
fluid \cite{Drozdowski2023}.

Here we extend this approach to a theory for
two species of NMII. Motivated by experiments by Weissenbruch et al. \cite{bastmeyer_graphic,eLife}, 
we incorporate phenomenological binding kinetics, including the competition of isoforms A and B for binding sites on actin filaments. 
This amounts to a volume exclusion interaction and results in an effective non-linear diffusion of myosins as known from the Tonks gas \cite{Tonks1932}. 
Our new model explains the experimentally observed distributions of NMIIA and NMIIB in a polarized migrating cell. 
Exploring the parameter space we can describe four different modes of cell motility: 
non-motile cells, decaying and stable pull-push velocity oscillations, and motile steady migration. Using both
numerical simulations and analytical theory, we
can obtain a complete phase diagram for our model.

The paper is organized as follows. We will start by defining the mathematical model based on active gel theory introducing the constitutive relation for stress. 
From binding kinetics we derive an effective nonlinear diffusion and demonstrate its consistency with the thermodynamics of a Tonks gas of two species. 
We then investigate the resulting closed boundary value problem and recapture experimental results in steady-state. 
Lifting the constraint of steady-state we demonstrate an oscillation mechanism for cell length and velocity. 
\red{Next} we introduce a limiting case which can be treated analytically and for which we present a phase diagram indicating four different modes of motility. 
\red{Lastly, we demonstrate the effect of actin polymerization.}

\section{Minimal Cell Model With Two Myosin Isoforms}

\subsection{Geometry and Mechanics of the cell}

Our one-dimensional model considers an infinitely compressible active gel.
As commonly assumed \cite{Bois_PhysRevLett2011, Recho_motility_initiation_and_arrest,Drozdowski2023}, the active contractile stress depends
linearly on the motor concentration $c$, 
$\sigma_{act}=-\chi c$, 
resulting in the constitutive relation
\begin{equation}
    \eta \partial_xv(x,t) = \sigma(x,t) - \chi_A c_A(x,t) - \chi_B c_B(x,t).
    \label{stress_relation}
\end{equation}
Here $v(x,t)$ is the velocity of the actin flow and 
$\eta$ the viscosity. 
The right hand side contains the total stress 
$\sigma(x,t)$ and the active stresses 
 caused by the 
myosin concentrations $c_A(x,t)$, $c_B(x,t)$ of the two isoforms. The minus signs relate to contraction
and the prefactors  $\chi_A$, $\chi_B$ 
quantify the contractility of the species.
Since the cell is adhering to the substrate, 
we assume homogeneous viscous 
drag and get the force balance
\begin{equation}\label{force_balance}
\partial_x \sigma = \xi v,    
\end{equation}
where $\xi$ is a friction coefficient. 
The cell occupies the range 
$x \in [l_-,l_+]$ and hence its length is given by $L(t)=l_+(t)-l_-(t)$ 
and its midpoint velocity by \red{$\Dot{G}(t)=\frac{d}{dt}(l_+(t)+l_-(t))/2$}. 
We use an elastic boundary condition,
$\sigma (l_\pm,t) = -k(L(t)-L_0)/L_0$, 
that models membrane tension and all other
restoring forces limiting the cell size 
\cite{Recho_motility_initiation_and_arrest,
DrozdowskiPRE2021,Drozdowski2023}.
Here $k$ is the effective spring constant
of the cell and $L_0$ its homeostatic length.

\subsection{Myosin Motor Binding Kinetics}

Myosin motors generate contractile forces
by binding to actin.
Informed by the myosin cross-bridge cycle \cite{cross_bridge_details1,cross_bridge_details2,erdmann2013stochastic,Erdmann2016}, we can formulate equations for their dynamic behavior. 
Although the cross-bridge cycle consists of multiple steps, we will solely distinguish between motors that are either attached to (index $a$) or detached from (index $d$) 
actin filaments to create an effective two-state model as considered in \cite{Recho_motility_initiation_and_arrest,STAM20151997,Drozdowski2023}.

In the unbound state, the motors can freely move inside the cytoplasm with diffusion 
constant $\Tilde{D}$ 
and bind to actin at a rate $k_{\text{on}}$. 
The binding process is subject to volume exclusion effects, i.e.~there must be a free spot available on the actin filament to be able to bind.
This is modeled by introducing a saturation
concentration $c_\infty$. When it is reached locally, 
no more binding is possible. 
The attached proteins move with the (retrograde) actin flow by advection with the velocity $v$.
Since the cytoplasm is 
considered as a reservoir for the motors, 
the detachment rate $k_\text{off}$ is assumed to be
independent of any interaction effects.
In general, the binding rates are specific for each isoform ($k_{\text{A},\text{on/off}}\neq k_{\text{B},\text{on/off}}$). In contrast, both the diffusion coefficient and the saturation concentration
are mostly determined by the size of the protein
which is the same for both isoforms.
This leads to the following two differential equations per isoform
\begin{subequations}
    \begin{align}
    \partial_t c_{i,a} &+ \partial_x (vc_{i,a}) = R(c_{i,a},c_{i,d}),
    \label{binding_kin1}
    \\
    \partial_t c_{i,d} &- \tilde{D} \partial^2_x c_{i,d} = -R(c_{i,a},c_{i,d}),
    \label{binding_kin2}
    \\
    R(c_{i,a},c_{i,d}) &= \frac{k_{i,\text{on}}}{c_\infty} (c_\infty - (c_{i,a}+c_{j,a})) c_{i,d} - k_{i,\text{off}}\, c_{i,a}.
    \label{reaction_kinetics}
    \end{align}
\end{subequations}
Here $R(c_{i,a},c_{i,d})$ is a nonlinear function 
describing the reaction kinetics and
$i,j$ refer to isoforms A and B, respectively. 
For the cellular behavior on the time scale of interest here, we do not include a production of myosin inside the cell, but rather require 
the total myosin concentration of each isoform to remain constant. 
This requirement then leads to no-flux boundary conditions for the two species, $\partial_x c_{\text{A,B}}(l_\pm,t)=0$. 

We now simplify the problem, 
as previously proposed for similar models \cite{Recho_motility_initiation_and_arrest,Drozdowski2023}. 
First, we assume
local chemical equilibrium, $R(c_a,c_d)=0$, 
and obtain expressions for the detached motors
\begin{align}
    c_{i,d}= \frac{1}{K_i} \cdot \frac{c_{i,a}c_\infty}{c_\infty-(c_{i,a}+c_{j,a})}
    \label{c_id},
\end{align}
where $K_i=k_{i,\text{on}}/k_{i,\text{off}}$
is the dissociation constant. 
Second, we take the limit of fast binding and fast diffusion ($K_\text{A}, \Tilde{D} \rightarrow \infty$) with $\Tilde{D}/K_\text{A} \rightarrow D_\text{A}$ and $\tilde{D}/K_\text{B} \rightarrow D_\text{A} K_r$, where we introduced $K_r=K_\text{A}/K_\text{B}$ as the ratio of the dissociation
constants of the two isoforms. 
Adding now Eq.~(\ref{binding_kin1}) and Eq.~(\ref{binding_kin2}) 
and inserting Eq.~(\ref{c_id}), 
we obtain one advection-diffusion equation 
for each isoform,
describing the effective dynamics of the 
bound myosin motors,
\begin{subequations}
\begin{align}
    \partial_t c_\text{A} &= - \partial_x (v c_\text{A}) + D_\text{A} \partial_x \mathcal{D}_\text{A}(c_\text{A},c_\text{B}),
    \\
    \partial_t c_\text{B} &= - \partial_x (v c_\text{B}) + D_\text{A} K_r \partial_x \mathcal{D}_\text{B}(c_\text{A},c_\text{B}).
\end{align}
\label{concentration_dyn}
\end{subequations}
Above, we have already dropped the indices $a$ and
in the following motors always refer to attached motors. 
The fact that the detached motor concentration drops out comes at the cost of concentration-dependent diffusion coefficients, which read
\begin{equation}
    \begin{split} 
    \mathcal{D}_i &= D_{ii} \partial_x c_i + D_{ij} \partial_x c_j 
    \\
    &=\frac{c_\infty(c_\infty-c_{j})}{(c_\infty-(c_\text{A}+c_\text{B}))^2} \partial_x c_{i} + \frac{c_\infty c_{i}}{(c_\infty - (c_\text{A}+c_\text{B}))^2} \partial_x c_{j}.
    \end{split}
    \label{nonlin_diff}
\end{equation}
The diffusion for both (actin-bound) myosin isoforms is non-linear in the concentration, 
and it shows cross-diffusion properties,
i.e.~a gradient of isoform $B$ influences the 
dynamics of isoform $A$ and vice versa. 
Note that for $c_\text{A}=c_\text{B}= c/2$, 
we obtain the model for a single motor species interacting through volume exclusion \cite{Drozdowski2023}. 
The concentration dependence 
in the prefactors of the gradients in  Eq.~(\ref{nonlin_diff})
arise from the  excluded volume interaction, 
quantified by the saturation concentration $c_\infty$. 
For concentrations approaching saturation the diffusion diverges, whereas for $c_\text{A},c_\text{B} \ll c_\infty$ we recover concentration-independent diffusion constants 
and vanishing cross-coupling.

Both nonlinear diffusion 
and cross-diffusion 
have been introduced and analyzed for 
other systems,
the former for instance for bacterial growth
\cite{muller2002morphological} and crowded
motor protein systems \cite{chelly2022cell,Drozdowski2023},
the latter in micelle solutions \cite{Leaist_micelles},
protein-polymer mixtures
\cite{Vergara_cross}
and reaction-diffusion systems
\cite{Vanag2009}.
We here intend to investigate their effects
in the context of intracellular flow, 
internal cell organization, 
cell polarization and cell motility.

\subsection{Myosin as Tonks gas}
\label{appendix_Tonks}

The nonlinear, concentration-dependent diffusion of Eq.~(\ref{nonlin_diff}) can also be derived  using arguments from classical irreversible thermodynamics.
Assuming a homogeneous 
"gas" of two species $\text{A, B}$ of hard spheres 
interacting only sterically (Tonks gas \cite{Tonks1932}), 
one finds a free energy density of
\begin{equation}
    \begin{split}
        f(c_\text{A},&c_\text{B}) = RT \Bigg[ c_\text{A} \ln \left( \frac{c_\text{A}}{c_\infty-c_\text{A}-c_\text{B}} \right) \\
        &+ c_\text{B} \ln \left( \frac{c_\text{B}}{c_\infty-c_\text{A}-c_\text{B}} \right)
        + c_\infty \ln \left( 1- \frac{c_\text{A}+c_\text{B}}{c_\infty} \right) \Bigg].
    \end{split}
    \label{free_energy_DFT}
\end{equation}
Here $R$ is the universal gas constant
and $T$ temperature.
This expression can be derived  by either using a density functional theory ansatz \cite{DFT_intro_tarazona,0d_cavitites,generalized_excess_energy,Lafuente_any_lattice} or statistical arguments similar to Flory-Huggins theory \cite{deGennes_flory_huggins,dill_bromberg_flory_huggins}.
 
The first two terms 
in Eq.~(\ref{free_energy_DFT})
arise from a Tonks gas with two different species of spheres.
However, the last term takes its form 
due to the introduction of holes into the system. 
These holes effectively constitute a third species.
Besides, the cytosol and actin network are considered as an additional dense solvent background, whose influence can be neglected due to the myosin size.
The saturation concentration then reads $c_\infty=c_\text{A}+c_\text{B}+c_\text{holes}$. 
In the dilute limit, i.e.~for large $c_\infty$, the last term reduces to the sum of $c_\text{A}$ and $c_\text{B}$ which resembles the free energy of the combination of two individual Tonks gases. 
Hence, the last term can be interpreted 
as the correlation between both species at concentrations of an order of magnitude similar to the saturation.

Within nonequilibrium thermodynamics,
fluxes are driven by gradients in 
conjugate thermodynamic forces.
The relevant couple in our case 
is particle flux 
and chemical potential.
\red{We use standard linear theory to express particle fluxes as a linear expansion of all thermodynamic forces, introducing the phenomenological Onsager coefficients $L_{ij}$ that in turn can also depend on the thermodynamic variables. Doing so we can relate the particle flux to the} diffusion within the multicomponent mixture
\begin{equation}
    \frac{1}{RT}\sum_{j \in \{\text{A,B}\}} L_{ij} \left( -\frac{\partial \mu_j}{\partial x} \right) = - \sum_{j \in \{\text{A,B}\} } D_{ij} \left( \frac{\partial c_j}{\partial x} \right),
\end{equation}
where $\mu$ is the chemical potential, 
$L_{ij}$ is the matrix of 
phenomenological coefficients 
and $D_{ij}$ the diffusion matrix.
Obtaining the chemical potential 
from the free energy via
$\mu_j = \partial f / \partial c_j$,
we get the same diffusion coefficients
as derived from binding kinetics, Eq.~(\ref{nonlin_diff}),
if the matrix of phenomenological coefficients is
\begin{equation}
    \mathbf{L}=\frac{c_\infty}{c_\infty -c_\text{A}-c_\text{B}} \begin{pmatrix}
        c_\text{A} & 0  \\ 0 & c_\text{B}
    \end{pmatrix}.
    \label{onsager_matrix}
\end{equation}

The concentration-dependent prefactor 
of Eq.~(\ref{onsager_matrix})
arises due to the excluded volume interaction. 
It is necessary in order for the steric repulsion not to vanish in the limiting case of one of the species'  concentration vanishing. 
Note that a species is not only affected by the excluded volume with itself, 
but also with the other species. 
For the dilute limit, which implies a large saturation concentration, 
the prefactor converges to unity
and one recovers the diagonal matrix $L_{ij}=c_i \delta_{ij}$ as typically used for the diffusion of different  species without explicit volume exclusion interactions, cf.~e.g.~Ref.~\cite{Esposito_2023}.

\subsection{Full Boundary Value Problem}

By combining the myosin dynamics Eq.~(\ref{concentration_dyn}), with diffusion coefficients given by Eq.~(\ref{nonlin_diff}), 
with the constitutive relation, Eq.~(\ref{stress_relation}), 
and the force balance,
Eq.~(\ref{force_balance}),
we can formulate the full boundary value problem. 

We introduce dimensionless variables and rescale length and position by the cell length $L_0$, 
time by $L_0^2/D_\text{A}$ and 
contractile stress by the spring constant $k$. 
The concentrations are normalized using the initial average concentration of isoform A, 
i.e.~$c_\text{A}^0=\int c_\text{A} dx/L_0$. 
The initial average concentration of B can then
be calculated using the relative abundance $c_r = c_\text{A}^0/c_\text{B}^0$. Furthermore, 
we introduce $\mathcal{L}= \sqrt{\eta/(\xi L_0^2)}$ as the ratio of the viscous and the \red{frictional} length scales \cite{mayerNature2010}, the dimensionless contractility $P_i=\chi_i c_\text{A}^0/k$ of the two isoforms
and the Péclet number $\Peclet=k/(\xi D_\text{A})$. 

We then map the problem into internal coordinates $u=(x-l_-)/L \in [0,1]$ to work on a fixed domain. The dimensionless boundary value problem then reads
\begin{subequations}
    \begin{align}
        -&\frac{\mathcal{L}^2}{L^2} \partial_u^2 \tilde{\sigma} + \tilde{\sigma} = P_\text{A} \tilde{c}_\text{A} + P_\text{B} \tilde{c}_\text{B},
        \\
        \partial_t \tilde{c}_\text{A} &= -\frac{1}{L}\partial_u (\tilde{v}\tilde{c}_\text{A}) + \frac{1}{L^2} \partial_u \mathcal{D}_\text{A}(\tilde{c}_\text{A}/L,\tilde{c}_\text{B}/L),
        \\
        \partial_t \tilde{c}_\text{B} &= -\frac{1}{L}\partial_u (\tilde{v}\tilde{c}_\text{B}) + \frac{K_r}{L^2} \partial_u \mathcal{D}_\text{B}(\tilde{c}_\text{A}/L,\tilde{c}_\text{B}/L).
    \end{align}
    \label{full_BVP}
\end{subequations}
Here $\tilde{c}_i(u,t) = L(t)c_i(u,t)$ and $\tilde{\sigma}(u,t)=L(t) \sigma(u,t)$ are rescaled concentrations and stress.
Both myosin isoform are advected by the actin retrograde flow with 
$\tilde{v}=\frac{\Peclet}{L^2}\partial_u \tilde{\sigma}-\dot{L}\left( u-\frac{1}{2}\right)-\dot{G}$. 
The elastic boundary condition reduces to $\tilde{\sigma}(u_\pm,t)=-L(t)(L(t)-1)$ with $u_\pm \in \{0,1\}$. 
The no-flux condition for the myosins
is given by 
$\partial_u c_{\text{A,B}}(u_\pm,t)=0$. 
The system of equations presented in Eq.~(\ref{full_BVP}) will now be analyzed
in the following
by continuation methods and direct numerical simulation. The main parameters are
$\Peclet$, $P_A$, $P_B$, $K_r$, $c_\infty$, $c_r$. 
The meaning of $\mathcal{L}$ has been discussed before in
\cite{DrozdowskiPRE2021,Drozdowski2023}.

\section{Results}

\subsection{Comparison to One-Species Models} 

As a background for the new two-species model,
we briefly recall the basic features of the 
one-species model. As studied in \cite{RechoPRL,Recho_motility_initiation_and_arrest}, the model for one species and a constant diffusion coefficient allows for a supercritical pitchfork bifurcation from sessile to moving 
cells as a function of $\Peclet$.
\red{The instability is due to the motors being
advected to the back by retrograde flow, from where they pull
further motors backward in a positive feedback
loop. Because the motor distribution at the back can become unrealistically high in this model, 
excluded volume effects have been introduced,
which imply nonlinear diffusion 
and smoothen the motor distribution \cite{Drozdowski2023}.
It was also shown that attractive interactions, modeling
myosin minifilament assembly, can render the
bifurcation subcritical, thus
leading to bistability and the possibility of
optogenetic switching. More recently,
it has been shown in a mathematically rigorous
analysis that this subcritical
bifurcation persists in a 2D version
of this model \cite{berlyand2025change}.}

Here we are motivated by the experiments by 
Weissenbruch et al.~\cite{eLife,bastmeyer_graphic}, who
have investigated the distributions of 
the isoforms NMIIA and NMIIB in 
polarized U2OS cells moving 
on a substrate in a steady fashion. 
As shown in Fig.~\ref{fig:fig1},
the more abundant, faster 
isoform A accumulates at the 
leading edge while the
stronger, higher duty ratio NMIIB 
at the trailing edge. Therefore we now have
generalized this model to two species.
For the moment being, we did not include
any attraction yet, thus our reference 
case is the Tonks gas and not the van 
der Waals fluid.

\subsection{Model Parameters} 

Informed by experimental data on crawling cells, we first
estimate the parameters of our model.
Due to $\xi \sim 2 \times 10^{14}\, \mathrm{Pa}\, \mathrm{s} \, \mathrm{m}^{-1}$ \cite{Drozdowski2023,barnhart_adhesion-dependent_2011}, $\eta \sim 10^5 \mathrm{Pa}\, \mathrm{s}$ \cite{Recho_motility_initiation_and_arrest,barnhart_adhesion-dependent_2011}, $k \sim 10^4\, \mathrm{Pa}$ \cite{barnhart_bipedal_2010,loosley_stick-slip_2012,eLife} and $L_0 \sim 20\, \mu \mathrm{m}$ \cite{Recho_motility_initiation_and_arrest} 
we obtain a (squared) relative viscous \red{length} scale of $\mathcal{L}^2=1.25$. Using $D_\text{A} \sim 0.1 \times 10^{-12}\, \mathrm{m}^2\, \mathrm{s}^{-1}$ \cite{uehara_determinants_2010,luo_understanding_2012,kolega_gradients_1993,chen_interplay_2023}, 
we infer the order of magnitude of the Péclet number to be $\Peclet \sim 100$. 

The isoform-specific (off-)binding rates determine $K_r$, which in turn serves as a measure of relative diffusion. 
While the binding rates of both isoforms are equal ($k_{\text{A,on}} = k_{\text{B,on}}=0.2\, \mathrm{s}^{-1}$), 
off-binding rates are $k_{\text{A,off}}=1.71\, \mathrm{s}^{-1}$ and $k_{\text{B,off}}=0.35\, \mathrm{s}^{-1}$ \cite{STAM20151997, Erdmann2016, parameters2, parameters1} leading to a relative diffusion of $K_r \approx 0.2$.  
The contractility parameters, as defined earlier, are of the order $P_i \sim 0.1$ \cite{Recho_motility_initiation_and_arrest}. 
Due to the different duty ratios of the isoforms A and B we find $P_\text{B} \approx 3.6 \cdot P_\text{A}$ \cite{Erdmann2016} and, thus, 
choose $P_\text{A} = 0.05$ and $P_\text{B} = 0.18$. Lastly, following arguments from Ref.~\cite{Drozdowski2023} 
we use $c_\infty = 4$ for the saturation concentration. The concentration ratio of $c_r\approx 1.5$ is taken from experiments \cite{bastmeyer_graphic}.

\begin{figure}
    \centering
    \includegraphics[width=\linewidth]{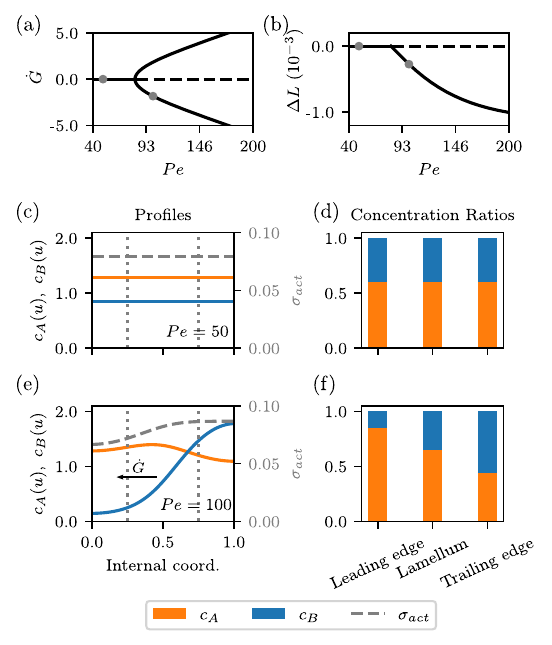}
    \caption{Simulation results for steadily moving cells. (a) Supercritical pitchfork bifurcation of cell velocity $\dot{G}$ as a function of the Péclet number $\Peclet$ (primary continuation parameter). 
    Solid lines mark stable, dashed lines unstable solutions. (b) Deviation of the cell length from its homeostatic value $\Delta L = L-L_0$ as a function of Péclet number. 
    (c) Myosin concentration profiles of the two isoforms and the resulting total active stress $\sigma_{act}$ of a sessile steady state.
    (d) The corresponding concentration ratios 
    subdivided into three compartments as indicated by the gray dotted lines in (c). (e) Concentration profiles and total active stress for a motile steady state (cell velocity direction to the left indicated with arrow)
    and the corresponding concentrations ratios in (f). 
    Parameters used: $\mathcal{L}^2=1.25, K_r=0.2, P_A=0.05, P_B=0.18, c_\infty=4,c_r=1.5$.
    }
    \label{fig:fig2}
\end{figure}

\subsection{Myosin Isoform Distributions in Steadily Moving Cells} \label{subsection:3a}

We solve the boundary value problem
for steady states 
using the numerical continuation software \textit{Auto07-p} \cite{auto_manual}.
The results for the parameters estimated above
are shown in Fig.~\ref{fig:fig2}. 
Panel~(a) and (b) show the bifurcation diagrams.
Below a critical value of $\Peclet_{crit} \approx 82$, the cell is in a sessile state with zero velocity (a)
and at its homeostatic length (b). 
The corresponding concentration profiles are shown in panel~(c): they are flat and hence symmetric for both isoforms. 
In (c) we also show the total active stress 
$\sigma_{act}(u)=-\chi_A c_A(u) - \chi_B c_B(u)$
as the dashed line,
which obviously is also flat.
We also define the concentration ratios as $c_i/(c_\text{A}+c_\text{B})$, with $i =\text{A,B}$.

At the critical Péclet number, a supercritical pitchfork bifurcation occurs, where the sessile state becomes unstable and stable motile steady states emerge. 
The two motile branches in panel~(a) are $\pm$-symmetric and belong to the same, slightly contracted branch of the cell's length in panel~(b). 
Whether one moves to the right or to the left
is determined by the initial conditions.
\red{Due to the high dimensionality of our nonlinear PDE system, 
analytical proof of attractor uniqueness is infeasible. 
However, Recho et al. \cite{Recho_motility_initiation_and_arrest} showed 
that in a similar single-species model, 
only the first supercritical pitchfork bifurcation leads to stable modes. 
Given this similarity, we expect the same here. 
Numerical simulations from varied initial conditions revealed no evidence of additional stable branches.}

In the steadily moving cell state beyond the bifurcation,
the myosin distributions are asymmetric, 
as depicted in panel~(e). This symmetry break in the concentrations induces
a polarized active stress profile that initiates cell polarization and motility. 
\red{In the shown example, the active stress is highest at the right edge ($u=1$), inducing a retrograde flow of actin to
this end. Therefore, the cell moves to the left, like in
the experimental examples from Fig.~\ref{fig:fig1}.}
The trailing edge is the locus of highest myosin concentration and, thus, highest active stress, in agreement to earlier theoretical works \cite{Recho_motility_initiation_and_arrest,DrozdowskiPRE2021} 
as well as experimental findings \cite{bastmeyer_graphic,weissenbruch2024actomyosin,eLife}.

As already discussed,
the two myosin isoforms 
and hence the concentration ratio 
are homogeneously distributed within the sessile state, see Fig.~\ref{fig:fig2} (d). 
Comparing to Fig.~\ref{fig:fig1} (d),
due to the lack of a polarization axis in the sessile state, we should identify the lamellum with the cell center and the region called "intermediate" region
in Fig.~\ref{fig:fig1} (d), and 
both edges with the periphery. 
In the motile steady state the stronger isoform B sits in the back, while the more diffusive isoform A is displaced towards the front causing the varying concentration ratios as shown in panel~(f). 
\red{This redistribution is due to
the balance of retrograde flow and diffusion, the former from flow towards the trailing edge and the latter effected by motors unbinding from actin.} 
The higher the diffusion, the less motors of the respective isoform arrive at the trailing edge. 
For the estimated value of $K_r=0.2$, isoform A has a larger effective diffusion than isoform B. 
Thus, B accumulates more quickly at the trailing edge. 
The effect of excluded volume then constrains A to accumulate further up front. 

\begin{figure}[t]
    \centering
    \includegraphics[width=\linewidth]{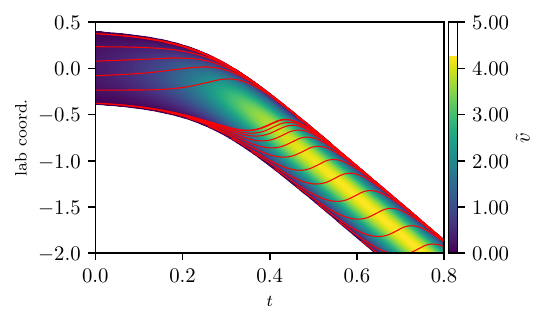}
    \caption{\red{Kymograph, i.e. trajectories in space and time, showing the advection through retrograde actin flow of cell that approaches steady-state motility. The flow lines (red) are directed towards the rear end and the flow velocity reaches a maximum value of $\tilde{v}_{\text{max}} \sim 4.0 \times 10^{-2} \mu \mathrm{m}\,\mathrm{s}^{-1}$.}}
    \label{fig:retrograde_flow_kymograph}
\end{figure}

\red{The symmetry breaking of concentration profiles is caused by the advection of myosin along the actin retrograde flow. The diffusion, depending on both $K_r$ and $c_\infty$, counteracts this by flattening the profiles. Hence, this polarization can only occur if the P\'eclet number is sufficiently large. 
This mechanism remains qualitatively unchanged when varying the other parameters. A more thorough sensitivity analysis of the parameters is performed in Appendix \ref{parameter_analysis}.
The theoretical results agree with the experimental data from Weissenbruch et al. \cite{bastmeyer_graphic}, compare Fig.~\ref{fig:fig2}(f) and Fig.~\ref{fig:fig1}(c).}

\begin{figure*}
    \centering
    \includegraphics[width=0.9\textwidth]{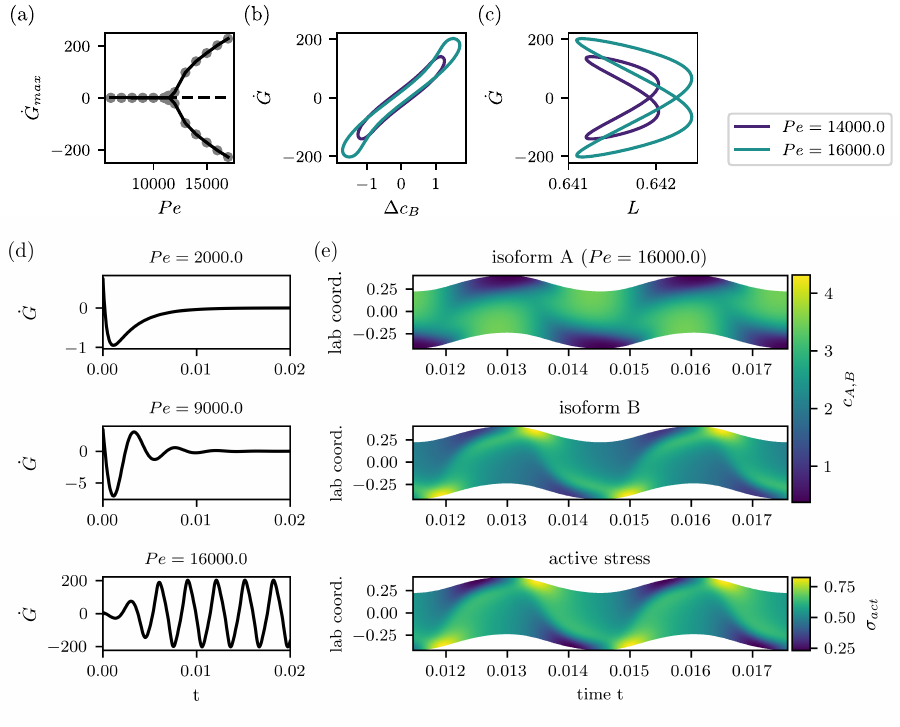}
    \caption{Oscillatory dynamics of the two isoform system. 
    (a) Hopf bifurcation at $\mathcal{K}_{crit}= 11500$ in the $\Dot{G}_{max}$-$\Peclet$-plane, where $\Dot{G}_{max}$ refers to the maximum velocity within an oscillation cycle. 
    (b) Exemplary limit cycles in the
    $\Dot{G}_{max}$-$\Delta c_B$-plane,
    where $\Delta c_B = c_B(l_+)-c_B(l_-)$ is a measure of the polarity of the myosin B distribution. The size of the limit cycle grows with $\Peclet$. 
    (c) Exemplary limit cycles in the plane cell velocity vs.~cell length. 
    (d) Examples of the three different
    motility modes 
    \,---\,sessile, transient oscillation and stable oscillation. Shown is
    the cell velocity over time. 
    (e) Kymographs visualizing the 
    pull-push-mechanism. Shown are the distributions of both myosin isoforms
    (upper and middle panel) and the resulting total active stress (lower panel) in a stable oscillatory state 
    at $Pe = 16000$.
    Parameters as in Fig.~\ref{fig:fig2},
    except for concentration ratio of $c_r=1$ and strongly increased dissociation/diffusion ratio $K_r=10$.
    }
    \label{fig:fig3}
\end{figure*}

\red{The myosin isoforms are mobile within the cell and are transported with actin filaments in the retrograde actin flow. 
Using finite volume simulations we can trace this flow $\tilde{v}$, introduced in Eq. (\ref{full_BVP}).
The result can be seen in Fig. \ref{fig:retrograde_flow_kymograph} as a kymograph,
showing a cell that, after an initial perturbation, approaches a state of steady motility to the left.
The retrograde flow is directed towards the rear end of the cell with a typical speed of $\tilde{v}_{\text{max}} \sim 10^{-2} \mu \mathrm{m}\, \mathrm{s}^{-1}$.
This value is small, but not unrealistic
\cite{MOGILNER2020143}.}

\subsection{Exploring Parameter Space and Occurrence of Oscillations}

We now lift the constraint of steady states,
solving the full boundary value problem  Eq.~(\ref{full_BVP}) numerically, 
and explore the parameter space
beyond the values estimated from 
experiments.
\red{As in the polarization mechanism, 
the most interesting parameters to study are 
the ratio of the
dissociation constants, $K_r$, the contractilities $P_A$ and $P_B$, and the P\'eclet number. These parameters again  qualitatively determine the motility mode of the system. In the following, we will keep the contractility values, but consider a large $K_r=10$ instead of the low $K_r=0.2$ from before.} In this parameter regime, we find 
additional modes of 
motility, in particular the 
possibility of oscillations, as shown in Fig.~\ref{fig:fig3}. 
For small P\'eclet number  
$\Peclet$ we again find stable sessile steady states. 
Increasing $\Peclet$
leads to the onset of transient oscillations, whose amplitudes decay 
towards the sessile steady state in the regime $ 6000 \lesssim Pe \lesssim 11500$. Beyond a critical value of 
$Pe_{\text{crit}} \approx 11500$, these oscillations prevail and become stable.
This behavior can be traced back to a Hopf bifurcation as shown in Fig.~\ref{fig:fig3}(a)-(c). 
Examples for the three different modes of motility\,---\,sessile, transient oscillation and stable oscillation\,---\,are shown in Fig.~\ref{fig:fig3}~(d).

For the oscillations to occur, the system requires a larger Péclet number. 
The increase of the critical Péclet number 
for the bifurcation as compared to Fig.~\ref{fig:fig2}
can be traced back
to the increase of the effective diffusion of B (by a factor of 50)
and the scaling of the Péclet number with the squared inverse of diffusion. 
The combination of the bifurcation diagram, cf.~panel~(a), and the growing closed limit cycles, cf.~panel~(b) and (c), demonstrate the emergence of a supercritical Hopf bifurcation towards stable oscillations. 
Compared to the cell velocity oscillations,
the oscillations in cell length
are smaller by several orders of magnitude,
see panel~(c). 

It is also interesting to study the  polarity, which can be quantified by 
the left-right edge difference 
of the stronger motor
$\Delta c_\text{B} = c_\text{B}(1)-c_\text{B}(0)$. The polarity and the velocity oscillate with the same frequency, 
see Fig.~\ref{fig:fig3}~(b).
while 
the length oscillates with twice the frequency of the velocity oscillation,
Fig.~\ref{fig:fig3}~(c).
This is because in each cycle the cell passes every length twice, once during contraction and once during elongation.

\begin{figure}[h!!!!]
    \centering
    \includegraphics[width=\linewidth]{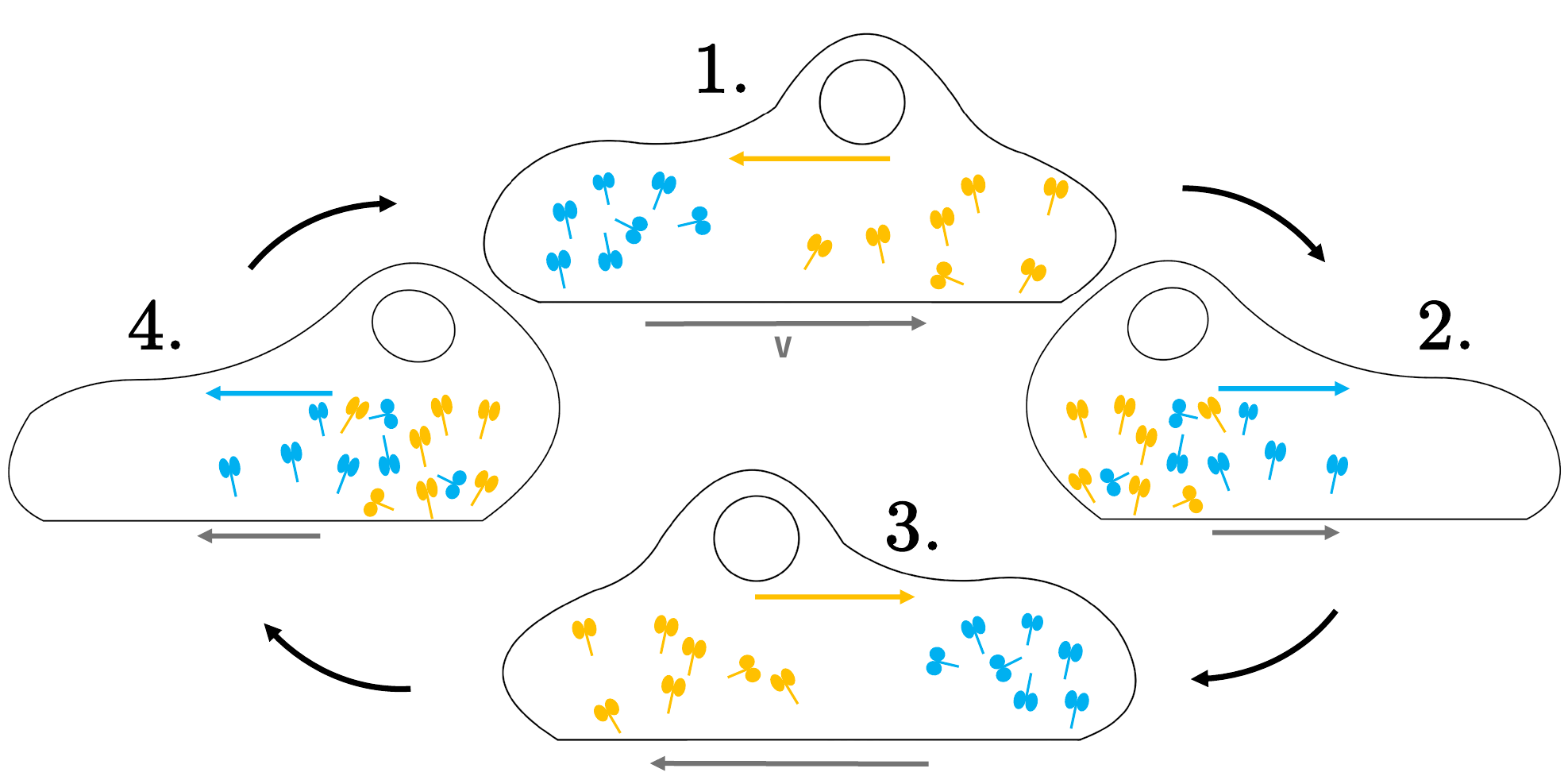}
    \caption{Schematic sketch of myosin dynamics within an oscillating cell. (1) The cell initially moves to the right. More contractile species B (blue) at the back causes net flow of A (yellow) to the left. (2) Then A displaces B. Net flow of B to the right (blue arrow). (3) Cell movement slows down and reverses. Cycle of "pull-and-push" starts in the other direction.}
    \label{fig:osci}
\end{figure}
How can the mechanism of the oscillation
be understood? \red{One has to note
that for $K_r >0$, one myosin isoform -- namely NMII B -- is more diffusive and, because of our choice of parameters here, also more contractile.} 
If we initially assume a spatial separation of both isoforms, the stronger isoform B then induces a larger\,---\,and hence dominating\,---\,contractile stress that induces an actin flow towards the position where isoform B is the most aggregated. 
Myosin motors of isoform A are advected along the retrograde actin flow.
As an effect of volume exclusion, isoform A subsequently displaces the more diffusive isoform B from its initial position and effectively "pushes" it to the other 
side of the cell. 
From this point on the pull-and-push cycle
repeats. A schematic sketch of the internal myosin dynamics during one oscillation cycle is depicted in Fig.~\ref{fig:osci}.

This pull-push mechanism is clearly visible within the simulation data
in Fig.~\ref{fig:fig3}(e),
showing kymographs (space-time-plots) 
of both isoforms (upper and middle panel) and the total active stress (lower panel)
driving the dynamics.
One can see that
both isoforms continuously exchange positions which in turn corresponds to an oscillatory active stress. 
Within the lab frame, the cell oscillates in space and velocity. 
As the oscillation in cell length
are much smaller than the one in cell velocity, it is barely visible in the kymographs. 

Increasing $\Peclet\gtrsim 17000$, our numerics, using a finite volume method implemented with the \textit{FiPy}-package for \textit{python} \cite{fipy-intro_article},
became numerically unstable. The reason is that at such high Péclet number, advection dominates the diffusion and causes a 
lack of regularization. A code better adapted to the hyperbolic nature of this regime could be developed.
We have rather chosen to circumvent this
numerical problem by considering
the rigid limit of the boundary value problem in the following, which also sheds
new light on the origin of the oscillation
and allows to establish a state diagram
for the system.

\subsection{The Rigid Limit and the State Diagram}

The much smaller oscillation amplitude of the cell length compared to the cell velocity, cf.~Fig.~\ref{fig:fig3}(c), 
suggests to look at the rigid limit of the problem, i.e.~where the cell has constant length, to simplify the analysis.

Similarly as suggested in Ref.~\cite{Recho_motility_initiation_and_arrest}, we introduce a new dimensionless stress $\sigma' = \sigma / (c_{\text{A}}^0\chi_A)$ leading to a new parameter for relative contractility $\chi_r = \chi_\text{A} / \chi_\text{B}$ and a new continuation parameter $\lambda = \Peclet P_\text{A}$. 
All other parameters from Eqs.~(\ref{full_BVP}) remain unchanged.  
The stiff limit then requires the contractility to be so small that the cell is not contracted, i.e.~$P_{\text{A,B}} \rightarrow 0, \ L \rightarrow 1$. This leads to the adjusted elastic boundary condition
\begin{equation}
    \sigma'(l_\pm, t) = \lim_{P_\text{A}\rightarrow 0} \lim_{L \rightarrow 1} \left( -\frac{1}{P_\text{A}} (L-1) \right) =: \sigma_0,
\end{equation}
where we assume the parameter $\sigma_0$ to be finite and implicitly defined through the remaining boundary conditions. 
We thus exchange the dynamics of the length, $L$, with the one from an implicitly defined "background stress" $\sigma_0$, 
which encodes the changes in the averaged level of contraction.

To simplify the problem we introduce the stress deviation field $s(u) = \lambda (\sigma'(u)-\sigma_0)$. In internal coordinates, the new boundary value problem now reads
\begin{subequations}
    \begin{align}
        \label{rigid_limit_s}
        &-\mathcal{L}^2 \partial_u^2 s + s + s_0 = \lambda \left( c_\text{A} + \frac{1}{\chi_r} c_\text{B} \right),
        \\[5pt]
        \partial_t c_\text{A} 
        &= \partial_u \left[ \left( \dot{G} - \partial_u s \right) c_\text{A} \right] 
        + \partial_u \mathcal{D}_\text{A}(c_\text{A},c_\text{B}),
        \\[5pt]
        \partial_t c_\text{B} 
        &= \partial_u \left[ \left(\dot{G} - \partial_u s \right) c_\text{B} \right] 
        + K_r \partial_u \mathcal{D}_\text{B}(c_\text{A},c_\text{B}),
    \end{align}
    \label{rigid_limit_PDE_system}
\end{subequations}
where $s_0=\lambda \sigma_0$ and the diffusion coefficients are the ones from from Eq.~(\ref{nonlin_diff}). 
The boundary conditions now read $\partial_u c_{\text{A,B}}(u_\pm,t)=s(u_\pm)=0$ and $\partial_u s(u_\pm) = \dot{l}_\pm \equiv \dot{G}$. 
Note that albeit having set the length to a constant value, which makes 
Eqs.~(\ref{rigid_limit_PDE_system}) much simpler than Eqs.~(\ref{full_BVP}),
we have not lost its degree of freedom, 
which is now encoded in the background stress $s_0$. 

We can now perform a linear stability analysis, i.e.~we expand all variables of our system in terms of small time- (and space-) dependent perturbations, added to the homogeneous steady state
\begin{subequations}
    \begin{align}
        s_0(t) &= s_0^0 + \delta s_0(t),\\
        s(u,t) &= s^0 + \delta s(u,t), \\
        c_i(u,t) &= c_i^0 + \delta c_i(u,t), \,\,\,i=\text{A,B}\,.
    \end{align}
\end{subequations}
We now neglect any dynamics of the background stress by assuming $\delta s_0(t)\equiv0$. From the stress boundary condition, we can infer $s^0=0$. 
The equilibrium values for the isoforms are chosen, as before, to be 
$c_\text{A}^0=1$ and $c_\text{B}^0=c_\text{A}^0 / c_r$.
When solving Eqs.~(\ref{rigid_limit_PDE_system})
numerically using the finite volume method, we however
keep the dependence on the background stress. It then has to be determined 
for every time step, such that all boundary conditions are satisfied.
In the analytical treatment, the perturbations are decomposed into Fourier modes 
of the form $\left[ X_i(q_i) \cos (q_i u) + Y_i(q_i) \sin (q_i u) \right] e^{\sigma_{q} t}$
that have to
satisfy the boundary conditions, 
leading to 
\begin{subequations}
    \begin{align}
        \delta s &= Y_\text{s}(2n\pi) \sin (2n\pi u) e^{\sigma_{q}t},\\
        \delta c_\text{A} &= X_\text{A}(n\pi) \cos ( n \pi u) e^{\sigma_{q}t}, \\
        \delta c_\text{B} &= X_\text{B}(n \pi) \cos (n\pi u) e^{\sigma_{q}t}.
    \end{align}
    \label{perturbations_fourier}
\end{subequations}
The spectrum of wave vectors  
is discrete with $n \in \mathbb{N}$ as we consider the finite domain $u \in [0,1]$.
Expanding the system of Eqs.~(\ref{rigid_limit_PDE_system}) to linear order in the perturbations and inserting Eqs.~(\ref{perturbations_fourier}) we obtain an eigenvalue problem 
$\sigma_q\,\mathbf{X}=\mathbf{M}_q\,\mathbf{X}$
for the vector
$\mathbf{X}=(\delta s,\delta c_\text{A},\delta c_\text{B})$,
with the matrix
\begin{equation}
    \mathbf{M}_q = 
    \left(\begin{smallmatrix}
        -4\mathcal{L}^2q^2-1 & \lambda & \lambda/\chi_r \\
        4c_\text{A}^0q^2 & -q^2 \frac{(c_\infty-c_\text{B}^0)}{\zeta_h} & -q^2 \frac{c_\text{A}^0}{\zeta_\text{h}} \\
        4c_\text{B}^0 q^2 & -q^2 K_r \frac{c_\text{B}^0}{\zeta_\text{h}} & -q^2 K_r \frac{(c_\infty-c_\text{A}^0)}{\zeta_\text{h}}.
    \end{smallmatrix}\right).
\end{equation}
In the submatrix of the myosin concentrations, we introduced the common factor $\zeta_\text{h}
=(c_\infty-c_\text{A}^0-c_\text{B}^0)^2/c_\infty$ coming from the nonlinear diffusion.
The construction of the matrix $\mathbf{M}_q$ through first and second order partial derivatives with respect to the perturbed variables renders the constant term $s_0^0$ irrelevant.

To predict the different modes of motility, we have to solve the eigenvalue problem. 
It is important to note that Eq.~(\ref{rigid_limit_s}) does not have any time derivatives, hence we take $\sigma_q=0$
in the first row of the 
characteristic equation.
This effectively sets one eigenvalue to zero and we are left with a pair of eigenvalues
that potentially are complex. 

\begin{figure*}
    \centering
    \includegraphics[width=0.9\textwidth]{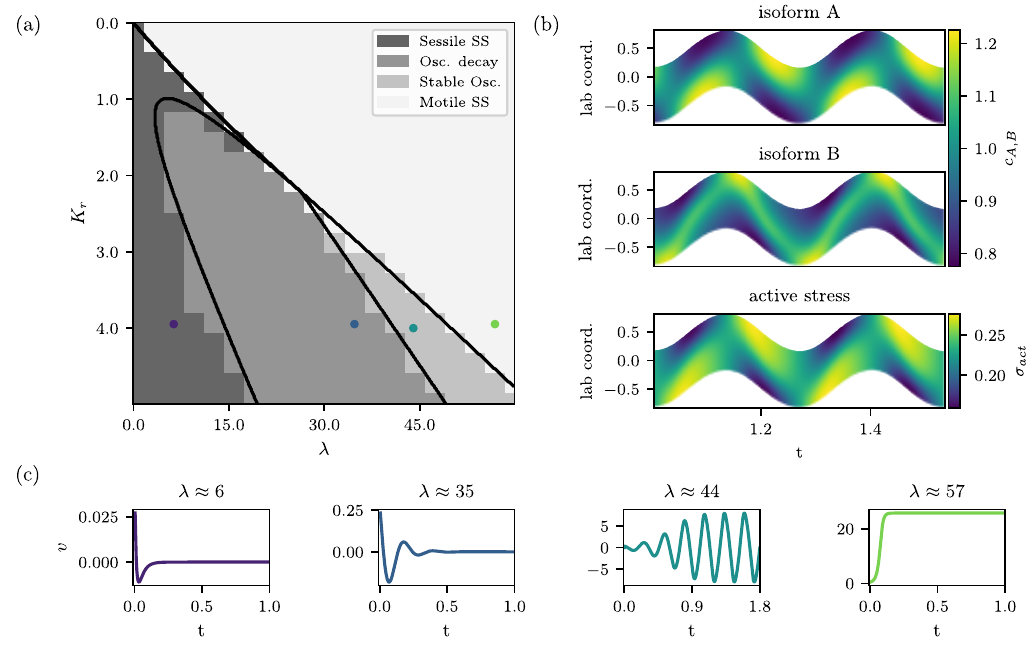}
    \caption{Motility modes in the rigid limit.  
    (a) State diagram of the four different motility modes depending on the relative diffusion $K_r$ vs.~the rescaled Péclet number $\lambda$. Numerical results obtained by finite volume simulations of equations (\ref{rigid_limit_PDE_system}) are coded in levels of gray. The curves are obtained from the linear stability eigenvalue problem and mark analytically predicted borders of the state diagram. 
    (b) Kymographs for the two motor 
    isoform concentrations and the resulting total active stress in a state of stable oscillations ($\lambda=44$). 
    (c) Examples of the four different motility modes as plots of velocity over time. $\lambda$ is varied while $K_r \approx 4$ was kept fixed. 
    Parameters correspond to the points indicated in (a).
    Parameters not mentioned have the same values as in Fig.~\ref{fig:fig3}.}
    \label{fig:fig4}
\end{figure*}

The results are shown in Fig.~\ref{fig:fig4}.
As shown in panel (a),
the numerical results of the full system of equations (\ref{rigid_limit_PDE_system})\,---\,shown as background colors\,---\,
agree very well with the linear stability analysis (curves). This  \textit{a posteriori} 
justifies the neglection of perturbations 
in $s_0$. We can identify four different modes of motility: sessile steady states, oscillations decaying towards a sessile steady state (transient oscillations), stable oscillations and motile steady states. 
Exemplary plots of 
velocity as a function of time for each 
of these modes are shown in panel~(c).

As also obvious from Fig.~\ref{fig:fig4}(a),
oscillations can only occur in case 
of $K_r > 1$. 
This implies $K_A > K_B$, meaning that the stronger isoform (myosin B) must have  larger effective diffusion than the weaker isoform (myosin A). 
In the case of stable oscillations, 
we recover the pull-push mechanism that includes the consecutive displacement among both isoforms as shown in the kymograph 
of Fig.~\ref{fig:fig4}(b).
This again proves the qualitative consistency of the rigid limit with the initial model, Eq.~(\ref{full_BVP}).

Note that in the linear analysis
we have only used the first mode ($n=1$), 
as it is the most dominant one for the dynamics \red{at the given hydrodynamic length of $\mathcal{L}\sim1.25$}:
in fact, it is the first antisymmetric mode that creates the strongest polarity 
for $\delta c_A$ and $\delta c_B$. 
Modes of even $n$ do not produce cell motility as they create symmetric concentration profiles. 
Higher modes of odd $n$ cause antisymmetric concentration profiles that consist of multiple maxima/minima, 
leading to overall less polarity. 
Hence, the first mode suffices for a qualitative analysis of the dynamics
of the systems, as also confirmed by the numerical solution. 
Also note that 
performing an analogous linear stability analysis for the original system Eq.~(\ref{full_BVP}) by simply fixing the length does not yield results consistent with the corresponding finite volume simulations.

\subsection{\red{Effect of Actin Polymerization}}

\begin{figure}
    \centering
    \includegraphics[width=\linewidth]{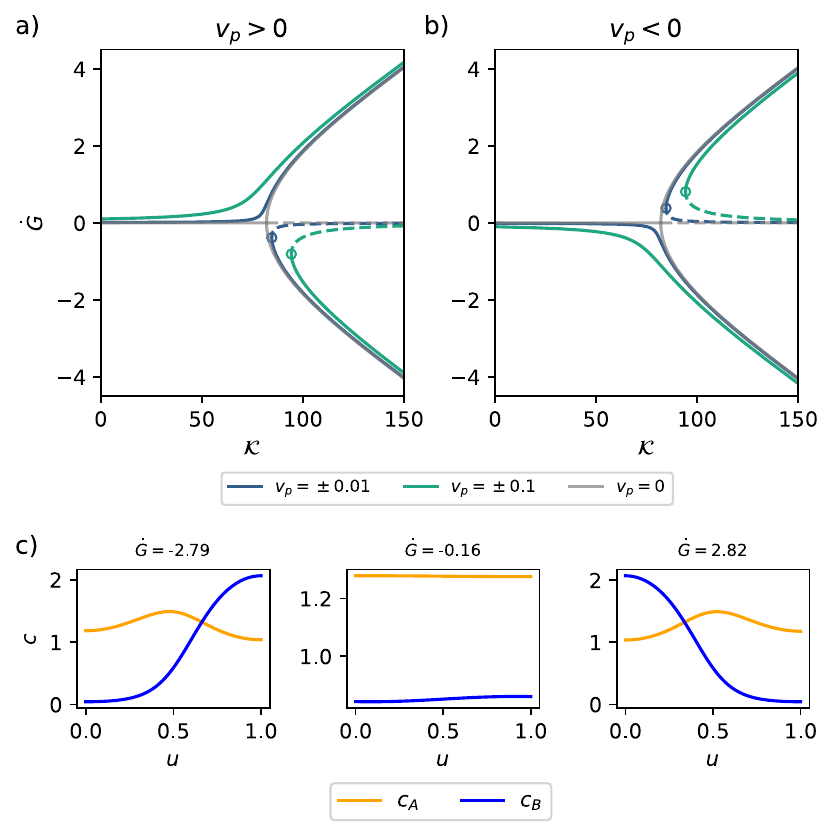}
    \caption{\red{Bifurcation diagrams for different polymerization velocities $v_p$. As soon as $v_p \neq 0$ the symmetry of the supercritical pitchfork bifurcation is broken and imperfect bifurcations with three branches emerge. (a) Bifurcation diagram for positive $v_p$. (b) Bifurcation diagrams for negative $v_p$. Encircled points mark changes in stability. (c) Concentration profiles taken at $\mathcal{K}=120$ for polymerization velocity $v_p=0.01$ at the stable solution branch, the unstable branch and the biased motile solutions (from left to right). Parameters used: $\mathcal{L}^2 = 1.25, K_r = 0.2, P_A = 0.05, P_B~=~0.18, c_\infty~=~4, c_r = 1.5$.}}
    \label{fig:polymerization_bifurcation}
\end{figure}

\red{Contraction-driven motility emerges by
an internal symmetry break as explained
above. Uneven actin polymerization at the cell
boundaries breaks front-rear symmetry
by construction and allows cells to 
quickly respond to external cues. For
one-species models, this effect
of actin polymerization has been demonstrated
before \cite{Drozdowski2023, Recho_motility_initiation_and_arrest, DrozdowskiPRE2021}.
Although here our focus is on contractility, for completeness
we have also performed a similar analysis for the two-species model presented here.

Polymerization can be incorporated into our model by modifying the force balance equation of Eq.~\ref{force_balance} adding the constant polymerization velocity $v_p$ to the contraction driven velocity $v(l_\pm)$ at the cell edges
\begin{equation}
    \dot{l}_\pm = \frac{1}{\xi}\partial_x \sigma(l_\pm(t),t) + v_p.
\end{equation}
This corresponds to symmetric polymerization where actin filaments are extended at the leading and retracted at the trailing edge with the same speed.
In order to keep the total mass of myosin constant, polymerization requires novel flux boundary conditions 
\begin{equation}
    \partial_x c_{A,B} (l_\pm,t) = -\frac{v_p}{D} \frac{\left(c_\infty-c_{A,B}(l_\pm,t)\right)^2}{c_{A,B}(l_\pm,t)}.
\end{equation}
This change poses difficulties when formulating the equations for steady-state motility as additional boundary terms are required. 
In the following analysis, we focus on the steady-state equations previously derived under no-flux boundary conditions. As a result, there is a slight loss of total myosin mass; however, this effect remains negligible for small polymerization velocities and over short time scales. While this approach is approximate, it nonetheless offers valuable qualitative insight into the system's behavior.

Fig.~\ref{fig:polymerization_bifurcation}(a) and (b) show the bifurcation analysis in steady state subject to small polymerization velocities. Any nonzero polymerization turns the pitchfork bifurcation into an imperfect bifurcation,
which means that it splits into a biased motile solution branch, 
a stable solution branch with opposite polymerization- and contraction-driven polarizations,
and an unstable branch unpolarized in the myosin distribution except for retrograde flow from polymerization.
This change in bifurcation type however neither changes the polarization mechanism from contraction 
nor the qualitative isoform distribution (cf. Fig.~\ref{fig:polymerization_bifurcation} (c)).}

\section{Discussion}

Motivated by the existence of several
isoforms of non-muscle myosin II (NMII), and
specifically by the experiments by Weissenbruch et al.~\cite{bastmeyer_graphic,eLife},
we have developed and analyzed a physical model 
including two interacting isoforms of NMII proteins. 
The interaction between the myosin isoforms A and B is mediated by excluded volume, that depends on the saturation concentration and leads to non-linear diffusion, derived both via a microscopic binding kinetics and non-equilibrium thermodynamics. 
The corresponding physics is that of a Tonks gas.
In the future, one could also include an attractive
interaction, thus changing to the van der Waals
fluid as a reference case, as suggested earlier for
a one-species active gel model \cite{Drozdowski2023}.
We expect that such a model extension would change
the bifurcation from supercritical to subcritical,
leading to bistability and the possibility of
optogenetic switching. Because this is not the focus
of the current work and would make analytical
progress even more difficult, here we here did not study this
potential extension. 

The two-species model introduced here recovers
the experimental distributions of the isoforms and in addition 
predicts the existence of a  variety of migratory modes. Regarding the steady states,
in the non-motile steady state both isoforms are homogeneously distributed and cause a symmetric active contractile stress. 
At sufficiently large Péclet numbers, however, this symmetry is broken and  polarized concentration fields and, consequently, polarized active stress profiles emerge.
This onset of polarization and migration
is due to a supercritical pitchfork bifurcation. 
For a steadily moving cell beyond the bifurcation, we find the
experimentally observed, isoform-specific myosin distribution profiles \cite{bastmeyer_graphic}:
the faster NMIIA accumulates at the leading edge,
while the stronger NMIIB localizes at the trailing edge.
\red{We also showed that our model can be extended to describe the effect of uneven actin
polymerization at the boundaries, which immediately stabilizes motile solutions
and leads to an imperfect pitchfork bifurcation, but otherwise does not 
change our conclusions.}

\red{Our one-dimensional model is motivated by the recent body of experimental work of cells migrating along lines and in channels. In order to analyze higher dimensions, other model classes such as phase field models are easier to implement \cite{Ziebert2012,Ziebert2016}, although
a recent mathematical analysis showed
how a rigorous bifurcation analysis can be
performed on a 2D model similar to the one described here \cite{berlyand2025change}.
Due to the similarities of this work with the one-dimensional one-species 
model \cite{DrozdowskiPRE2021}, we would also expect our two-species model to behave 
similarly in two dimensions.}

We have explored further modes of motility besides the steady state solutions. Oscillations occur in case that the more contractile isoform is also more diffusive (i.e. $P_B > P_A, \ K_r > 1$),
via a cyclic pull-push-mechanism. This mechanism works best for similar total abundances of both isoform, $c_r\sim1$. 
It has been suggested that different myosin isoforms can be regulated independently 
\cite{Bresnick_CurrOpCellBiol99_myosin_regulation_review, Barua_PNAS12_Actpmyosin_regulation_tropmyosin_isoforms, Clayton_CurrBiol10_Differential_regulation_myosin_yeast}, 
which would allow for the relative diffusion to be controlled via $K_r$. 
Albeit not observed in the \red{animal} cell type investigated in Refs.~\cite{eLife,bastmeyer_graphic}, 
it is not unlikely that these oscillations might occur
in other cell types \red{(e.g. from other
species)}. \red{Another interesting 
option is to genetically engineer cells
such that these oscillations occur, 
either spontaneously or triggered by
external control, in particular by optogenetics.
If oscillations can be controlled in this manner,
it also would be easy to distinguish them
from oscillations that occur as a result of
non-linear friction, as described by clutch models 
\cite{sens2020stick,Amiri2022.08.30.505377, Ron_PRR20_1d_cell_motility_patterns}.}

In the oscillatory state, 
the amplitude of the cell length oscillation
is much smaller than the one of velocity. 
This suggested to study the rigid limit, 
which enabled an analytic linear stability analysis and allowed to determine
the full state diagram of the system,
containing sessile cells, cells with transient oscillations and persistent oscillations, as well as steady motile cell solutions. This phase diagram is in good agreement with finite volume simulations of the full model. 

In summary, here we have shown that active gel
theory can be extended to also describe the 
effect of having two complementary isoforms
at work. The fact that we could derive the same
model from both kinetic and a thermodynamic 
perspectives demonstrate its fundamental 
nature and paves the way to also include other
biologically relevant effects into the model. 
In particular, a third major molecular constituent of
contractile structures in cells is $\alpha$-actinin,
which like NMII also exists in different isoforms
and which also competes for binding to the actin
filaments. \red{In our model framework, such a competitor
would increase excluded volume for the NMII 
and therefore decrease active stress. At the same
time, $\alpha$-actinin might increase elasticity
and therefore make the cell effectively stiffer. 
It would be interesting to include such effects,
but this also would introduce many more model parameters, that then
had to be estimated from experiment. The same
is true for replacing the simple friction by
a molecular clutch \cite{sens2020stick,Woessner_2024}.
Despite this complexity, however, the work presented here clearly shows
that molecular details can be fruitfully 
included in fundamental models for cell motility.}

\section{Acknowledgments}

We thank Martin Bastmeyer and his group for helpful discussions. 
O.M.D. acknowledges support by the Max Planck School Matter to Life supported by the German Federal Ministry of Education and Research (BMBF) in collaboration with the Max Planck Society.
We also acknowledge support by the Deutsche Forschungsgemeinschaft (DFG, German Research Foundation, project numbers EXC 2181/1 - 390978043,
EXC 2082/1-390761711 and SFB-1638/1 – 511488495 - P03).
U.S.S. is member of the Interdisciplinary Center for Scientific Computing (IWR) at Heidelberg. 

\appendix
\renewcommand{\thefigure}{S\arabic{figure}}
\setcounter{figure}{0}

\section{Parameter Analysis}
\label{parameter_analysis}

In the following, we analyze the system's dependence on several parameters and their effect on the dynamics. We use steady state solutions throughout. For all analyzed values we find supercritical pitchfork bifurcations at $\mathcal{K}_{crit}$ in analogy to what is shown in Fig.~\ref{fig:fig2}(a).

\begin{figure*}
    \centering
    \includegraphics[width=0.55\linewidth]{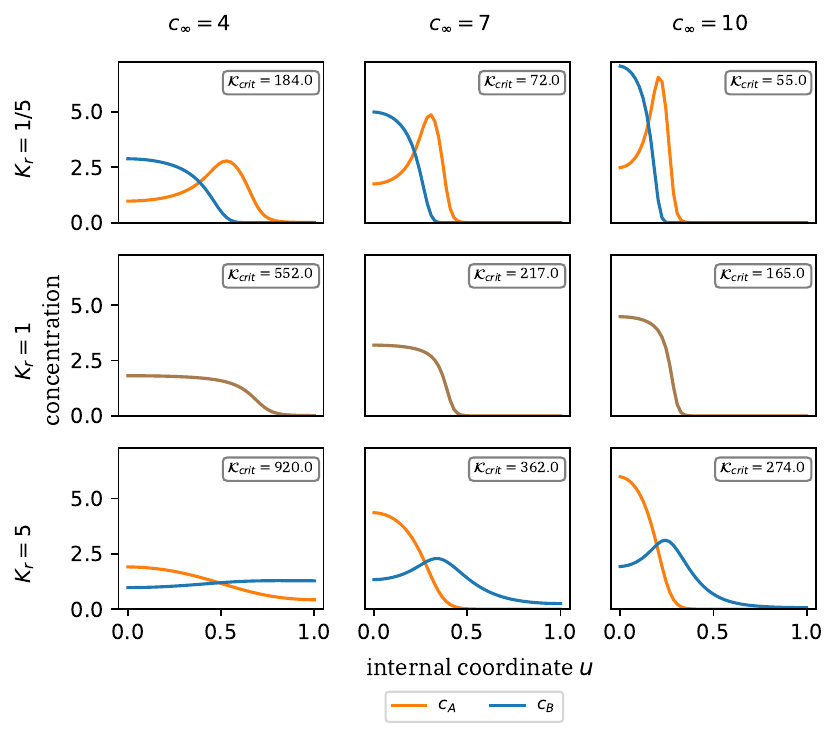}
    \caption{\red{Concentration profiles for different saturation concentrations $c_\infty$ and relative diffusivities $K_r$. All profiles taken at $\mathcal{K}=1000$, contractilities $P_A=P_B=0.06$ and total concentration ratio $c_r=1$. For $K_r=1$ both profiles are equal.}}
    \label{fig:Kr_cinf_9plot}
\end{figure*}

Fig.~\ref{fig:Kr_cinf_9plot} shows the effect of various combinations of parameter values of saturation concentration $c_\infty$ and relative diffusivity $K_r$. Following the rows, we see that a low saturation concentration flattens and widens the concentration profiles. In contrast, high saturation allows for sharper peaks and overall spatially narrower profiles inside the cell. This holds for all $K_r$. 
Larger saturation concentrations allow for stronger polarization at the edges. Hence, the bifurcation point $\mathcal{K}_{crit}$ marking the transition from sessile to motile states shifts to smaller values for increasing saturation concentration. 
The columns reveal the influence of relative diffusion. The isoform with the lowest diffusivity sits further in the back (species $B$ for $K_r<0$, species $A$ for $K_r>0$). We can explain this by the transport of myosin along the retrograde actin flow. 
The retrograde flow moves towards the trailing edge.
For the cases depicted here of positive cell velocity 
we find a flow towards the left. During advection, the motors also diffuse away and unbind from actin. 
The higher the diffusion, the less motors arrive at the trailing edge to initiate motility. This also explains why the critical Péclet number ($\mathcal{K}_{crit}$) increases with increasing $K_r$. 
According to Eq.~(\ref{full_BVP}), the diffusion of species $A$ is not affected by $K_r$, which only influences the diffusion of species $B$. 
Hence, an increase in $K_r$ leads to a higher total diffusion with less biased concentration and therefore less asymmetric active stress profiles, causing a later onset of motility. 
Once the space at the trailing edge is occupied by the less diffusive species, the other species can only place itself closer to the cell center due to the volume exclusion interaction.

\begin{figure*}
    \centering
    \includegraphics[width=.55\linewidth]{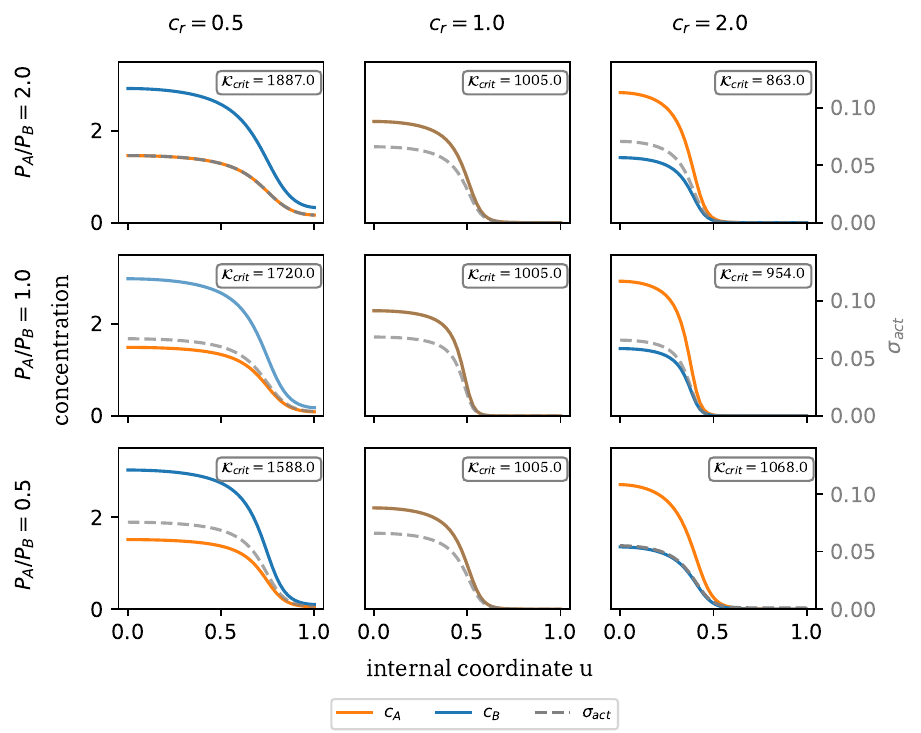}
    \caption{\red{Concentration profiles and active stress for varying relative contractility and concentration ratios. All profiles are plotted at $\mathcal{K}=2500$. Fixed parameters are $K_r=1.0, c_\infty=5$. The following values for the respective contractility were used: first row: $P_A=0.02, P_B=0.01$; second row: $P_A=P_B=0.015$; third row: $P_A=0.01, P_B=0.02$.}}
    \label{fig:Pr_cr_plot}
\end{figure*}

Fig.~\ref{fig:Pr_cr_plot} shows the effect of 
varying the ratio of total concentration and the ratio of contractility. Since $K_r$ is kept constant the concentration profiles of $A$ and $B$ have a similar qualitative shape and positions in each respective plot.
The relative magnitudes of the concentrations originate from $c_r$. For $c_\infty \neq 0$ one species is more abundant -- species $A$ when $c_r>1$, species $B$ when $c_r<1$ while for $c_r=1$, the profiles and the active stress $\sigma_{act}$ are also quantitatively equal. For all plots, the sum of both contractilities is equal.
For each row, we find the narrowest and sharpest concentration profiles in the last column at the maximum and the flattest and broadest profiles in the first column at the minimum value of $c_r$. The varying profile widths can be explained by the way that we have normalized the concentrations: in internal coordinates $c_{A}^0=1$ and $c_{B}^0= 1/c_r$. Hence, we find the highest sum of integrated concentrations for $c_r=0.5$ and the lowest for $c_r=2.0$ ($c_{A}^0+c_{B}^0=3, \text{ resp. } 1.5$). 
Consequently, two phenomena can be identified. 
First, at low $c_r$ (effective larger total concentration) the system is more significantly affected by the saturation concentration. 
The intensified repulsive volume exclusion force leads to flattening and broadening of the profiles. Second, for a larger total number of motors, 
increased myosin advection is required to facilitate polarization, as can also be seen in the shifting
bifurcation point $\mathcal{K}_{crit}$.
With an increase in $c_r$, a decreasing trend in $\mathcal{K}_{crit}$ is observed. An exception to this trend appears in the final plot of the last row. 
There, the combination of $A$ being more abundant while $B$ is more contractile leads to a higher critical Péclet number.
The influence of the contractility ratio on the shapes and magnitudes of the concentration is negligible.

\clearpage

%

\end{document}